\documentclass[]{aipproc}
\layoutstyle {8x11single}

\begin{document}

\setlength{\tabcolsep}{9mm}

\topmargin -21mm


\title
{Measuring CMB Polarization with ESA PLANCK SubMM-Wave Telescope}

\author{Vladimir Yurchenko$^{\ast \dag}$}{
  address={$^{\ast}$ Experimental Physics Department, 
National University of Ireland, Maynooth, Co. Kildare, Ireland},
  address={$^{\dag}$ Institute of Radiophysics and Electronics, 
National Acad. Sci., 12 Proskura St., Kharkov, 61085, Ukraine}
}


\begin{abstract}
We analyze the polarization properties of the tilted off-axis 
dual-reflector submillimeter-wave telescope on the ESA PLANCK Surveyor 
designed for measuring the temperature anisotropies and polarization 
characteristics of the cosmic microwave background.

\end{abstract} 

\maketitle

\section{Introduction} 

The dual-reflector submillimeter-wave telescope on the ESA PLANCK 
Surveyor is being designed for measuring the temperature 
anisotropies and polarization characteristics of the cosmic 
microwave background (CMB). 

Our research is concerned with one of its focal plane instruments, 
Far-IR High Frequency Instrument (HFI) [1], which will cover six 
frequency bands centered at $100$, $143$,$217$, $353$, $545$ and 
$857 GHz$ providing the sensitivity of $\Delta T/T \sim 10^{-6}$
and the angular resolution down to $5$ arcminutes. The HFI consists
of an array of $36$ horn antenna structures (Fig.~1, a) feeding 
the bolometric detectors which will be cryogenically cooled to a 
temperature of $100 mK$.

One objective of the research is the optimization of the HFI 
optical design and the computation of the HFI beam patterns. 
The challenge of the problem is that the telescope is electrically 
large ($D/\lambda=4300$ at $\lambda=350 \mu m$) and consists of 
two essentially defocused ellipsoidal reflectors providing a very 
large field of view at the focal plane.

Another objective is the characterization of the polarization 
properties of the multi-beam telescope system. The CMB polarization 
is expected to be at a level of only 10\% of the temperature 
anisotropy quadrupole, and the success of the measurements will 
depend crucially on the precise knowledge of the polarization 
properties of the telescope.

\section{Formulation of the problem}

While the performance of the antenna structures can be thoroughly 
tested in terrestrial conditions, the coupling of the HFI with the 
telescope is, basically, optimized through the computer simulations. 

Among various simulation techniques, physical optics (PO) is the 
most adequate one for the given purpose. However, conventional 
implementations of the technique [2, 3] do not fit the size of 
the problem. Commercially available packages are also very limited 
in their capacity to rigorously answer this sort of questions.
For example, even the best commercial software requires many 
hours of the full-scale physical optics computation of the main 
beam of the telescope at the relatively low frequency of $143 GHz$, 
while all conventional physical optics codes collapse at the 
highest frequencies of $545 - 857 GHz$.

To solve the problem, we developed a special PO code [1] that 
allowed us to overcome the limitations of a generic approach for 
large multi-reflector systems and perform typical simulations of 
the telescope in the order of minutes. Generally, it requires 
only $1$ minute for a polarized Gaussian beam at the frequency 
of $143 GHz$ and about $30$ minutes for the beam of $30$ modes 
at $857 GHz$ using a PC Pentium III ($500 MHz$) under the Linux 
operating system. 

%
\begin{figure}
(a)
    \includegraphics[height=.29\textheight]{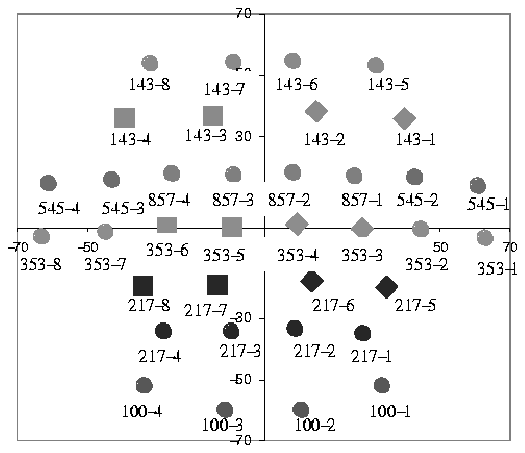}
(b)
    \includegraphics[height=.29\textheight]{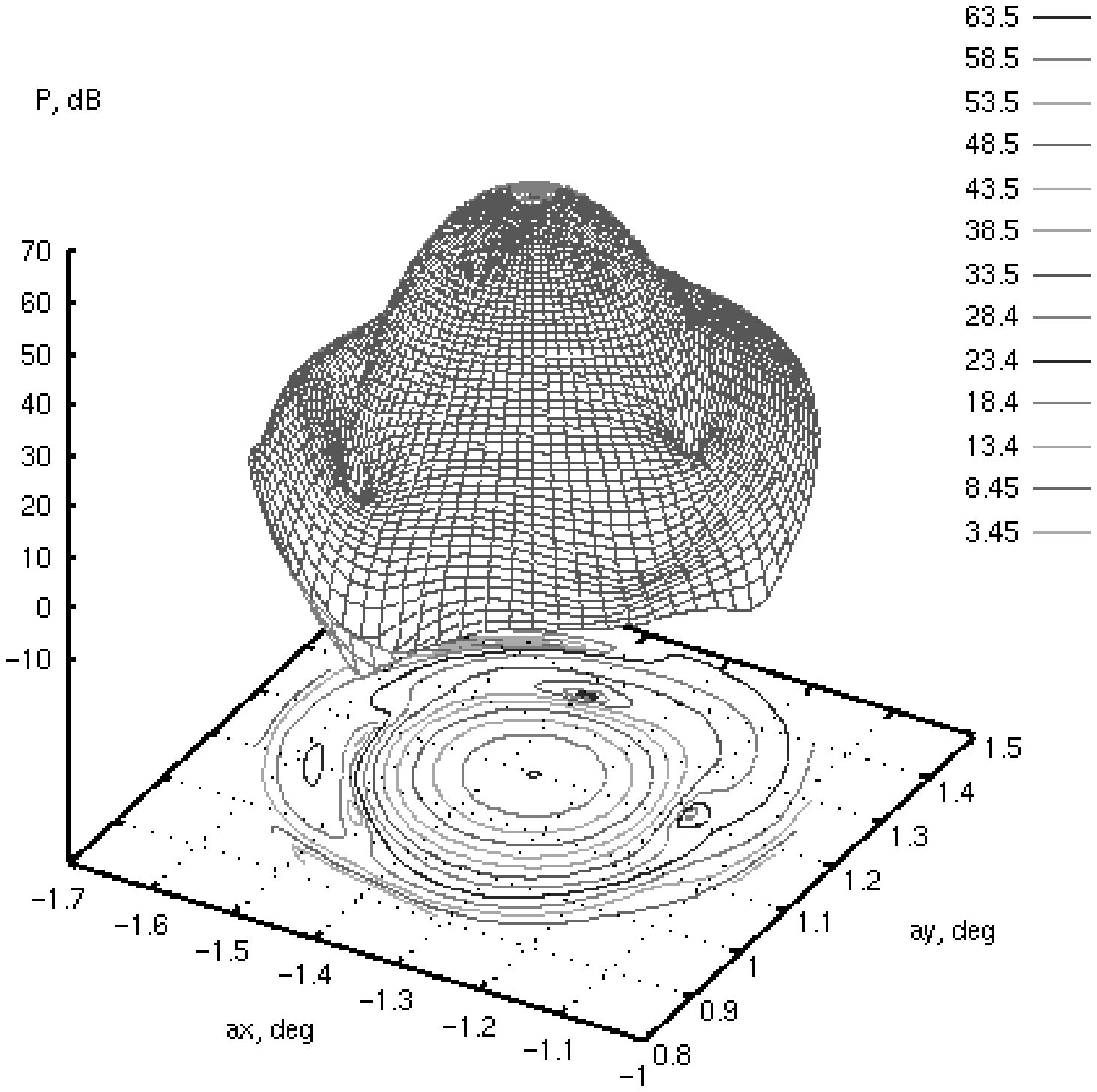}
 \caption{(a) Horn positions on the focal plane as seen from the 
telescope. (b) Total power of the telescope beam 143-1a computed 
for the clipped Gaussian feed model F3 with $a=3.9mm$, $w=2.8mm$, 
and $L=120mm$ ($f=143 GHz$).}
 \label{fig1}
\end{figure}
%
%
\begin{figure}
(a)
    \includegraphics[height=.3\textheight]{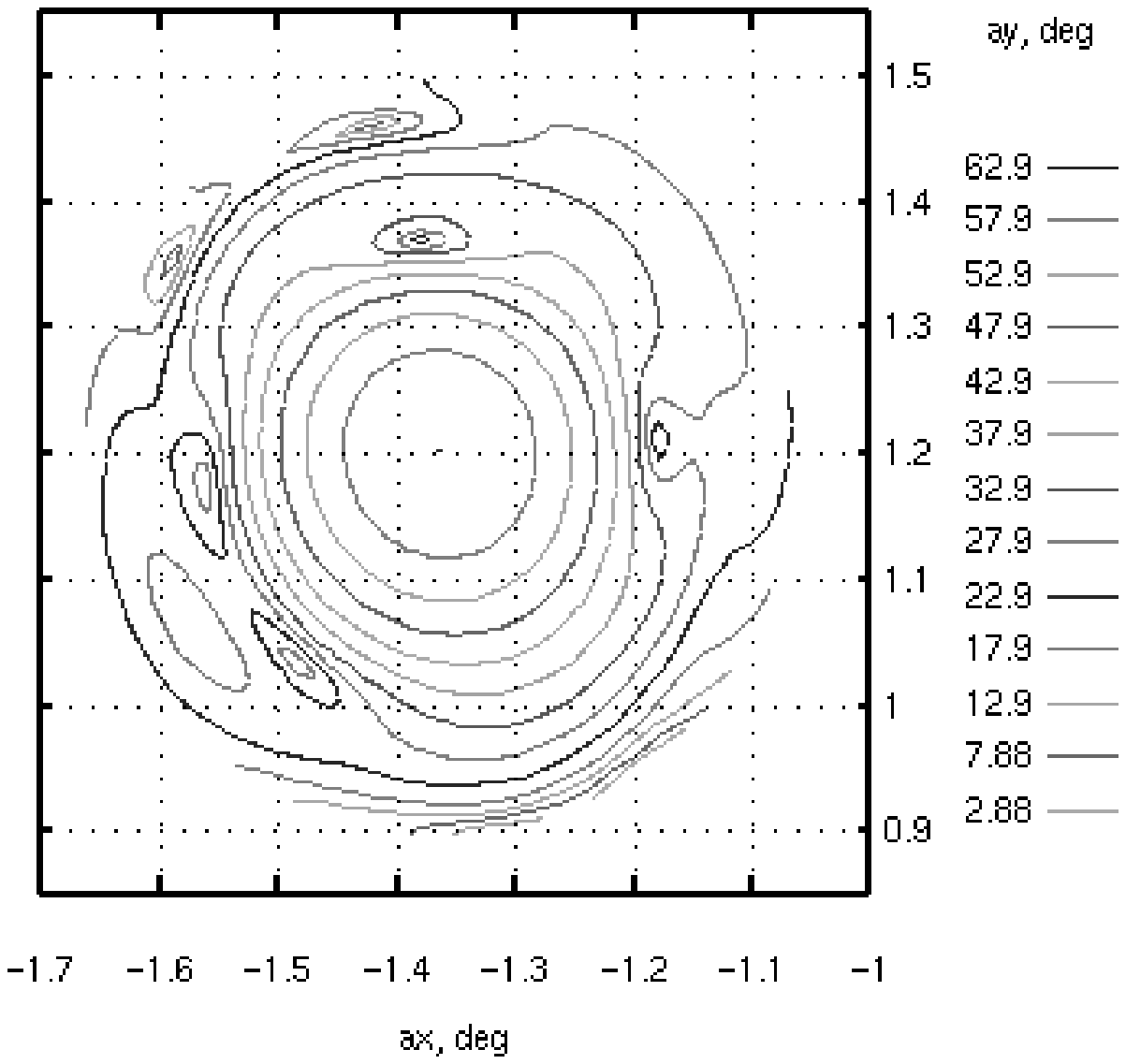}
(b)
    \includegraphics[height=.3\textheight]{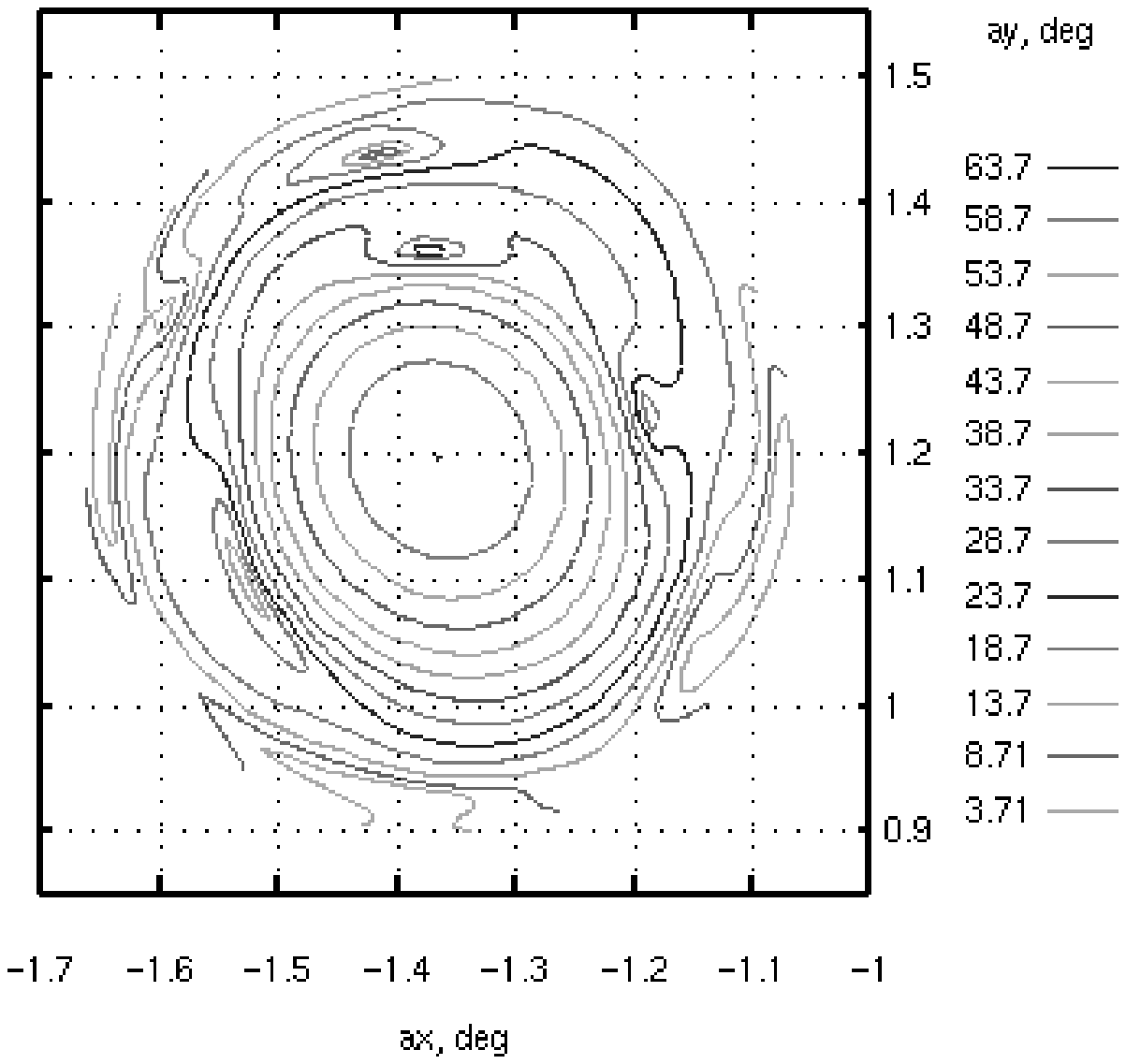}
 \caption{Power pattern of the telescope beam from the enhanced 
horn HFI-143-1a ($\psi_a =+45^{\circ}$, feed model F1) at the 
frequency 
(a) $121 GHz$ and (b) $166 GHz$.}
 \label{fig2}
 \end{figure}
%

We simulated the beams of twelve linearly polarized Gaussian horns 
operating in the transmitting mode at the nominal frequencies 
$143$, $217$, and $353 GHz$.
Each horn is designed for the simultaneous measurement of two 
orthogonal linear polarizations, 'a' and 'b', of the incoming 
radiation. The polarization is characterized by the tilt $\psi$
of the electric field at the beam axis in the sky with respect 
to the local vertical as defined below. The tilt is 
$\psi_a =+45^{\circ}$ and $\psi_b =-45^{\circ}$ for the horns 
143-1, 143-2, 217-5, 217-6, 353-3, and 353-4 (see Fig.~1, a), 
while $\psi_a =0^{\circ}$ and $\psi_b =90^{\circ}$ for the other 
six polarized horns (the angle $\psi$ is measured clockwise 
from the upward vertical direction as seen from the telescope).

We compared three different models of the horn feed, F1, F2 and 
F3, when the source field is specified in different manner. 
The feed model F1 is represented by the far-field power and phase 
patterns of the actual corrugated horn in $0^{\circ}/45^{\circ}
/90^{\circ}$ cuts which are used in a somewhat approximate manner 
for computing the incident complex-vector electromagnetic 
field on the secondary mirror. 
The feed F2 is specified by the 'clipped Gaussian' model 
distribution of the horn aperture electric field which is 
accurately propagated to the secondary mirror and further 
through the system. 
Finally, the feed F3 is represented by the far-field patterns 
similar to the model F1 which, however, are computed precisely 
from the clipped Gaussian aperture field of the model F2 and 
then used in an approximate manner identical to the feed model F1.

The electric field at the aperture of the Gaussian horn
in the feed models F2 and F3 is specified as
\begin{equation}
\vec E(r) = 
      E_0 \exp(-r^2/w_a^2) \exp(-iqr^2) \ \vec e, 
            \qquad  0\le r \le a  \\ 
\end{equation}
where $r$ is the radial coordinate, $w_a$ is the beam radius at the 
horn aperture, $a$ is the horn aperture radius,  $q=\pi/\lambda L $, 
$L$ is the curvature radius of the wavefront at the aperture 
(approximately, $L$ is the horn slant length), and 
$\vec e=\vec e_{a,b}$ is the unit polarization vector
($\vec E(r)=0$ if $r > a$). 
The aperture model of this kind is quite accurate in representing 
the field of real Gaussian horns. 
In particular, both the power and phase patterns of such a horn 
computed at the frequency $100 GHz$ coincide perfectly well with 
the experimental data available for the model Gaussian horn 
designed for this frequency. However, some discrepancies in the 
sidelobes at the level of about $-30 dB$ appear for the horns 
specifically optimized for the higher angular resolution.

The comparison of the feed models F2 and F3 shows that the minor 
differences can only be observed in the polarization patterns 
at the periphery of the telescope beam while the power patterns 
of the main beam are identical for both models. The beams computed 
with the feed models F1 and F3 are also very similar when the 
parameters of the model F3 are properly adjusted. In this case, 
also, it is only the polarization at the periphery of the beam 
that differs slightly for different models while the power 
patterns are very similar even at the level below $-30 dB$.  
Finally, the orientation of the polarization vector on the 
beam axis in the sky field is precisely the same for all the 
feed models, and the polarization pattern, in general, is 
rather independent of both the horn pattern and of the fine 
features of the propagation model used in the simulations.
It proves that the far-field model F1 can safely be used for the 
simulation of the telescope beams from the real horns despite 
the approximations used in the model.


\section{Polarization of the Gaussian beams}

Fig.~1, b, shows the power pattern of the telescope beam H-143-1 
as projected on the plane normal to the telescope line-of-sight 
at the $(0,0)$ point ($a_x$ and $a_y$ are the horizontal and 
vertical axes on the plane, respectively, measured in degrees). 
The beam axis is at the point $a_x= -1.3645^{\circ}$, 
$a_y= 1.1975^{\circ}$ which is defined as the point of maximum 
power of the beam. The pattern is computed for the clipped Gaussian 
feed model F3 with adjusted parameters $a=3.9mm$, $w=2.8mm$, and 
$L=120mm$, simulating the enhanced horn operating at the 
frequency $f=143 GHz$.

The beam is well shaped down to almost $-30dB$ below the maximum 
and can be approximated by a Gaussian function (at this level, 
the pattern does not depend on polarization) with the full beam 
width of $W_{min}=7.05$ arcmin and $W_{max}=7.50$ arcmin 
measured at $-3dB$.
The polarization of the beam is generally elliptical except 
precisely at the beam axis where it remains linear. In order 
to achieve the required orientation of the polarization pattern 
in the sky, we should orient the polarization vector $\vec e$ 
properly on the horn aperture~[1]. 

For immediate comparison of polarizations measured by different 
horns when scanning through the sky, we should use easily aligned 
directions in the sky as equivalent reference polarization axes 
for different beams. Such directions are the meridians in the 
spherical frame of the telescope, with the pole being the
telescope spin axis tilted by $\eta =85^{\circ}$ with respect
to the line-of-sight (the meridians define local verticals at
various observation points while the parallels are local
horizontals that constitute the orthogonal directions).

Now, we should define the reference axis for the polarization
vector $\vec e$ of a tilted horn. We define the reference axis
as the direction of $\vec e$ in the horn aperture plane that is
projected on the vertical axis of the focal plane.
The orientation of $\vec e$ is specified by the angle $\phi$ 
in the horn aperture plane measured from this reference in a 
clockwise direction when looking from the horn to the secondary 
mirror. 

Using these definitions, we have found that the beam of the 
H-143-1 horn is polarized at its axis at the required angles 
of (a) $\psi_a = +45^{\circ}$ and (b) $\psi_b = -45^{\circ}$ 
if the horn polarization vector $\vec e$ is specified by the 
angle $\phi_a=+42.99^{\circ}$ and $\phi_b=-47.01^{\circ}$, 
respectively. 

When the polarization vector $\vec e$ is properly oriented, 
the cross-polarized component of the far field measured along  
the respective orthogonal direction in the sky is minimized.
In such a case, the power pattern of the cross-polarized 
component is, basically, determined by the power pattern 
of the semi-minor axis of the polarization ellipse at each 
observation point of the beam. This pattern is very much 
the same for any orientation of the polarization vector and 
can be approximated as follows 
\begin{equation}
P_{cr}=P_{cr0} \ [(\theta/\theta_0) \ \sin(\varphi-\varphi_0)]^p \ 
\exp[(\theta/\theta_0)^q]
\end{equation}
where $P_{cr0}, \ dB$, is the maximum power of the minor axis 
component achieved at the points specified by the polar angle 
$\theta_0$ and the azimuthal angles $\varphi_0 \pm 90^{\circ}$ 
measured from the center of the pattern which is located at the 
beam axis (the values of the parameters in this approximation 
depend on the position of the horn considered but the magnitude 
of $P_{cr0}$ is, typically, more than $30 dB$ below the maximum 
value of the total power of the beam).

\begin{figure}
(a)
    \includegraphics[height=.3\textheight]{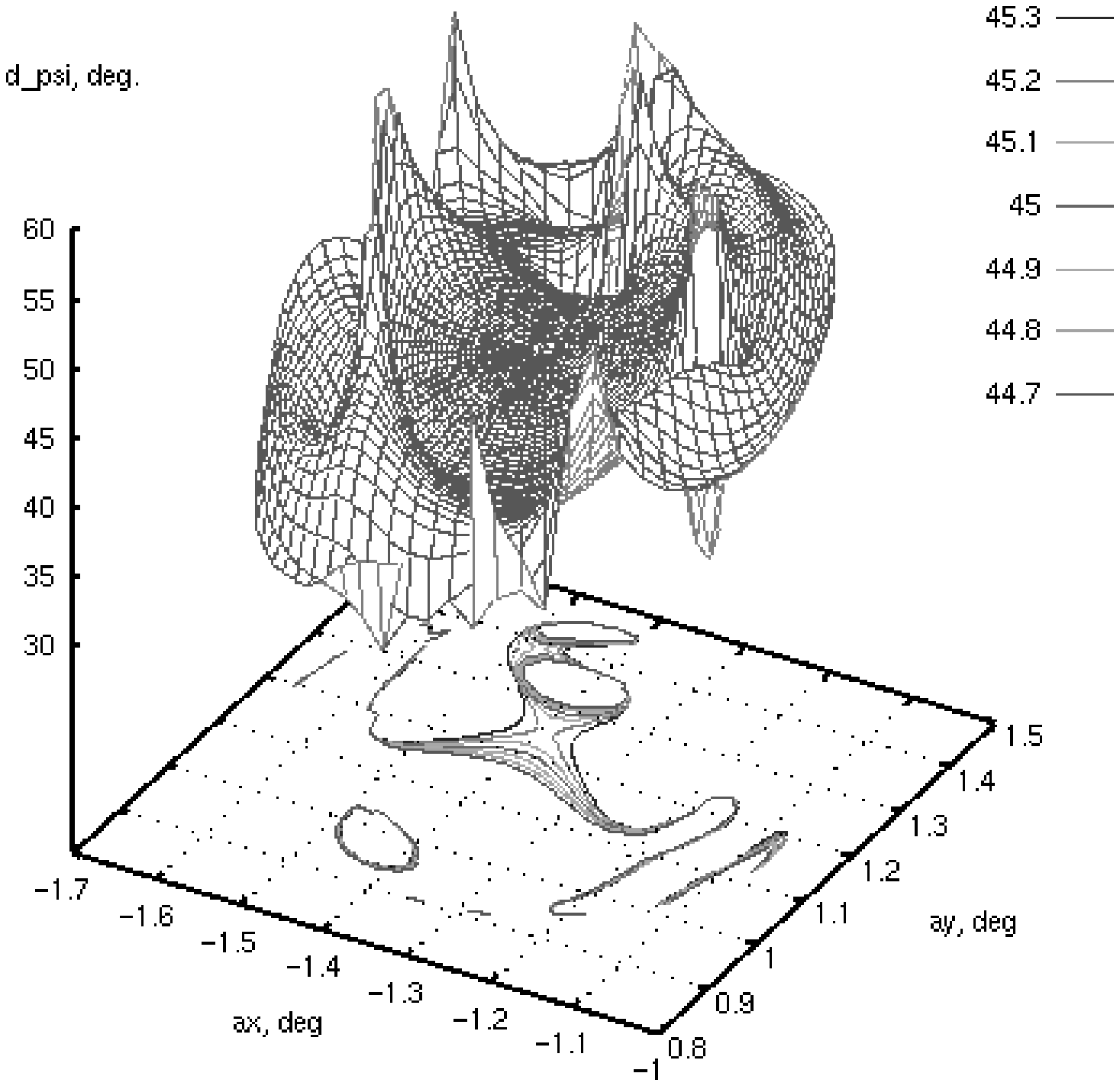}
(b)
    \includegraphics[height=.3\textheight]{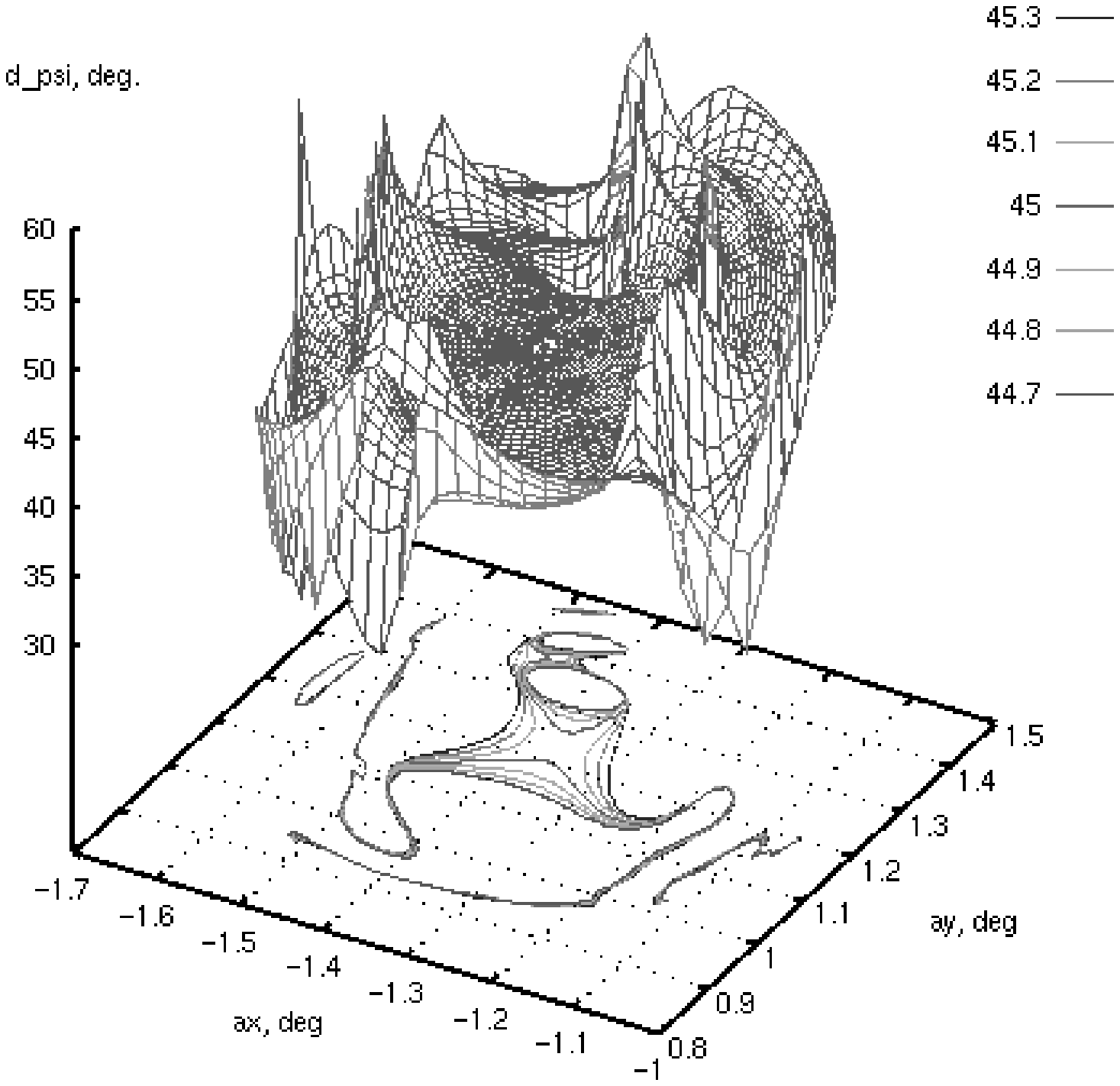}
 \caption{Deviation of the major axis of polarization ellipse 
from local vertical in the beam from the enhanced horn HFI-143-1a 
($\psi_a =+45^{\circ}$, feed model F1) at the frequency 
(a) $121 GHz$ and (b) $166 GHz$.}
 \label{fig3}
 \end{figure}
%
 \begin{figure}
(a)
    \includegraphics[height=.3\textheight]{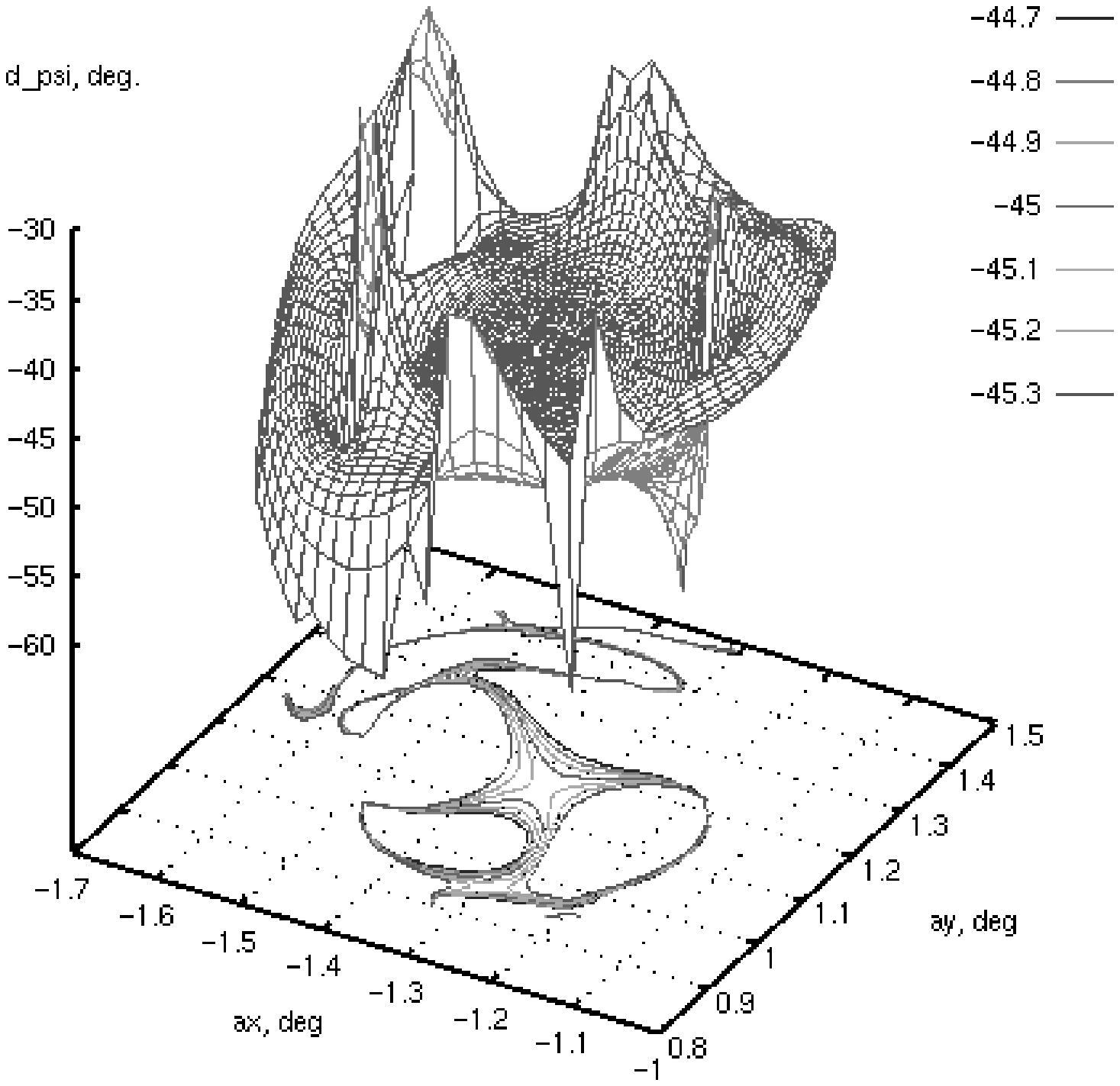}
(b)
    \includegraphics[height=.3\textheight]{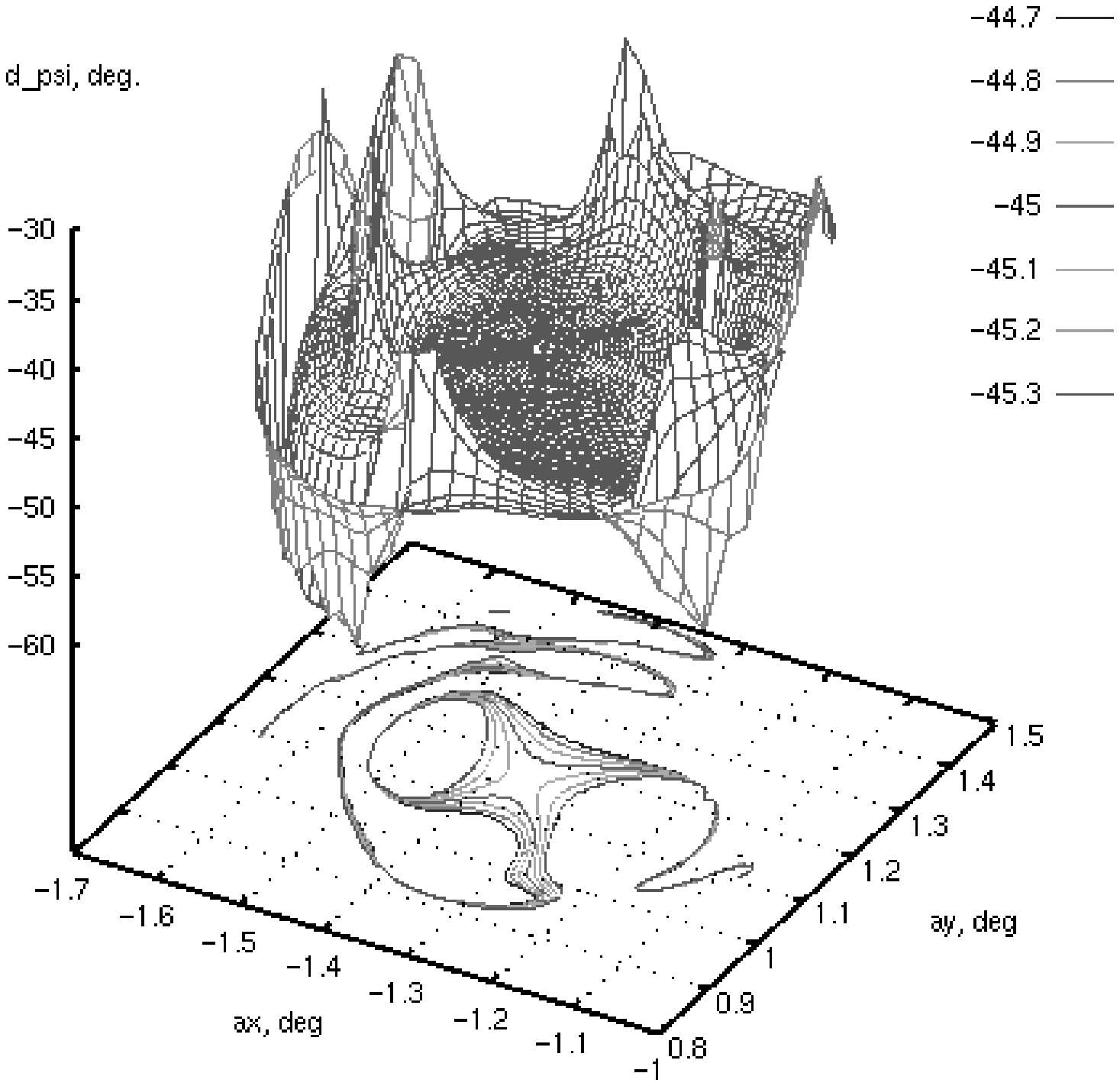}
 \caption{Deviation of the major axis of polarization ellipse 
from local vertical in the beam from the enhanced horn HFI-143-1b 
($\psi_b =-45^{\circ}$, feed model F1) at the frequency 
(a) $121 GHz$ and (b) $166 GHz$}
 \label{fig4}
 \end{figure}

Since the actual frequency bands of the horns are rather wide 
($121-166$, $182-252$, and $295-411 GHz$), the far-field patterns 
of real horns differ essentially at the lower and upper 
frequencies of each band. It results in somewhat different power 
and polarization patterns of the telescope beam as shown in 
Figures 2 - 4. Nevertheless, despite these differences, the 
polarization angles at the beam axis are the same irrelevant of 
the horn patterns and of the operating frequency. 

The values of the aperture polarization angle $\phi$ computed for 
all the HFI horns as well as the direction cosines (the Cartesian 
components of the unit vector) of the aperture electric field with 
respect to the coordinate frame $M1$ of the primary mirror are 
summarized in the Table 1.

\begin{table}
\caption{\bf Orientation of the Polarization Vector $\vec e$ 
of the Aperture Electric Field}
\begin{tabular}{cr@{.}lr@{.}lr@{.}lr@{.}l}
\hline \\ [-2mm]
\multicolumn{1}{c}{\bf HFI horn} & 
\multicolumn{2}{l}{\bf Aperture Angle} & 
\multicolumn{6}{c}{\bf \hspace{-5mm} 
Unit Vector of the Aperture Electric Field} \\
\multicolumn{1}{c}{} & 
\multicolumn{2}{c}{ $\phi$ [degrees] } & 
\multicolumn{2}{c}{\bf $e_{X_{M1}}$ } & 
\multicolumn{2}{c}{\bf $e_{Y_{M1}}$ } & 
\multicolumn{2}{c}{\bf $e_{Z_{M1}}$ } \\ [1mm] 
\hline \\ [-1mm] 
143-1-a & \hspace{5mm}  42&99 & 0&629015 &  0&679549 &  0&377562 \\
143-1-b & \hspace{5mm} -47&01 & 0&521115 & -0&728982 &  0&443874 \\ 
143-2-a & \hspace{5mm}  44&17 & 0&597142 &  0&696388 &  0&398076 \\
143-2-b & \hspace{5mm} -45&83 & 0&552563 & -0&716862 &  0&425186 \\ 
143-3-a & \hspace{5mm}   0&82 & 0&813844 &  0&014303 &  0&580907 \\
143-3-b & \hspace{5mm} -89&18 & 0&031357 & -0&999321 & -0&019324 \\ 
143-4-a & \hspace{5mm}   2&10 & 0&812137 &  0&036506 &  0&582323 \\
143-4-b & \hspace{5mm} -87&90 & 0&080095 & -0&995568 & -0&049291 \\ [2mm] 
217-5-a & \hspace{5mm}  42&99 & 0&661010 &  0&680077 &  0&317114 \\
217-5-b & \hspace{5mm} -47&01 & 0&567397 & -0&729548 &  0&381863 \\ 
217-6-a & \hspace{5mm}  44&11 & 0&634880 &  0&695677 &  0&336097 \\
217-6-b & \hspace{5mm} -45&89 & 0&593249 & -0&717633 &  0&364772 \\ 
217-7-a & \hspace{5mm}   0&89 & 0&869188 &  0&015525 &  0&494238 \\
217-7-b & \hspace{5mm} -89&11 & 0&029397 & -0&999362 & -0&020308 \\ 
217-8-a & \hspace{5mm}   2&01 & 0&867847 &  0&034982 &  0&495598 \\
217-8-b & \hspace{5mm} -87&99 & 0&066276 & -0&996754 & -0&045699 \\ [2mm] 
353-3-a & \hspace{5mm}  43&70 & 0&634926 &  0&690048 &  0&347422 \\
353-3-b & \hspace{5mm} -46&30 & 0&571370 & -0&722094 &  0&390020 \\ 
353-4-a & \hspace{5mm}  44&70 & 0&610581 &  0&703339 &  0&364012 \\
353-4-b & \hspace{5mm} -45&30 & 0&594981 & -0&710743 &  0&375289 \\ 
353-5-a & \hspace{5mm}   0&58 & 0&853206 &  0&010120 &  0&521476 \\
353-5-b & \hspace{5mm} -89&42 & 0&020092 & -0&999707 & -0&013472 \\ 
353-6-a & \hspace{5mm}   1&51 & 0&852069 &  0&026308 &  0&522767 \\
353-6-b & \hspace{5mm} -88&49 & 0&052455 & -0&998000 & -0&035275 \\ [2mm]
\hline
\end{tabular}                           
\end{table}

\section{Conclusions}

A fast physical optics code has been developed for the 
analysis of the dual-reflector submillimeter-wave telescope 
on the ESA PLANCK Surveyor. The code overcomes the limitations 
of a generic approach for large multi-reflector quasi-optical 
systems and can perform typical simulations of the telescope 
in the order of minutes. 

Simulations of the telescope beams from the linearly polarized 
horns have shown that the far-field of the telescope is, generally, 
elliptically polarized except precisely at the beam axis where 
the linear polarization is preserved. 
The magnitude of the minor semi-axis of the polarization ellipse 
in the telescope beam remains at the level of $-30dB$ below 
the maximum total power of the beam even for the most tilted 
horns located at the edge of the HFI horn array.

When rotating the polarization vector of the horn field about 
the horn axis, the major axes of the polarization ellipses 
at the central part of the telescope beam rotate virtually 
by the same angle about the beam axis while the power patterns 
of both the co- and cross-polarized components of the far field 
remain basically unchanged. 

Thus, the orthogonal polarization directions at the beam axis 
in the sky are transformed into the orthogonal directions in the 
focal plane of the telescope. 
Some deviations from this basic rule which occur mainly 
at the periphery of the beam are of minor importance since they
virtually do not contribute to the measured total power of the 
polarized component.

\section{Acknowledgments}

The author is grateful to Bruno Maffei for the power and phase 
patterns of enhanced horns and to Yuying Longval for providing 
the updated positions and aiming angles of the HFI horns along 
with the map of the focal plane. The author would like to 
acknowledge the support of Enterprise Ireland.

\end{document}